\documentclass[fleqn,10pt]{wlscirep}
\usepackage[utf8]{inputenc}
\usepackage[T1]{fontenc}
	\usepackage{makecell}

\title{Dynamic relationship between XRP price and correlation tensor spectra of the transaction network}
\author[1,2,*]{Abhijit Chakraborty}
\author[2,+]{Tetsuo Hatsuda}
\author[1,$\dag$]{Yuichi Ikeda}

\affil[1]{ Kyoto University, Graduate School of Advanced Integrated Studies in Human Survivability, Kyoto, 606-8306, Japan}
\affil[2]{RIKEN Interdisciplinary Theoretical and Mathematical Sciences Program, Saitama, 351-0198, Japan}

\affil[*]{chakraborty.abhijit.7y@kyoto-u.ac.jp}
\affil[+]{thatsuda@riken.jp}
\affil[$\dag$]{ikeda.yuichi.2w@kyoto-u.ac.jp}

\begin{abstract}
The emergence of cryptoassets has sparked a paradigm shift in the world of finance and investment, ushering in a new era of digital assets with profound implications for the future of currency and asset management.
A recent study showed that during the bubble period around the year, $2018$, the price of cryptoasset, XRP has a strong anti correlation with the largest singular values of the correlation tensors obtained from the weekly XRP transaction networks.  In this study, we provide a detailed analysis of the method of correlation tensor spectra for XRP transaction networks. We calculate and compare the distribution of the largest singular values of the correlation tensor using the random matrix theory with the largest singular values of the empirical correlation tensor. 
We investigate the correlation between the XRP price and the largest singular values for a period spanning two years. We also uncover the distinct dependence between XRP price and the singular values for bubble and non-bubble periods. The significance of time evolution of singular values is shown by comparison with the evolution of singular values of the reshuffled correlation tensor. Furthermore, we identify a set of driver nodes in the transaction networks that drives the market during the bubble period using the singular vectors. 
\end{abstract}
\begin{document}

\flushbottom
\maketitle
%
%
\section*{Introduction}
Cryptoasset has emerged as a new asset class that has gained enormous popularity and attention from investors worldwide. The rapid growth and widespread adoption of cryptoassets have led to a surge in prices and market capitalization recently. However, high volatility in the cryptoasset market has led to concerns about its stability and reliability as an investment.
One of the major challenges in the cryptoasset market is the presence of anomalies in price movements. An anomaly refers to an observation that deviates significantly from the expected or normal pattern. Anomalies in the cryptoasset market can occur due to various factors, including market manipulation, insider trading, regulatory changes, or technical issues. Cryptoasset uses a decentralized digital ledger called a blockchain to store transaction data. The blockchain is a distributed ledger that contains a record of every transaction that has ever occurred on the network. Each block in the blockchain contains a set of transactions, and these blocks are linked together in a chronological chain. The blockchain provides a secure and transparent way to store transaction data without the need for a central authority or intermediary.

 Cryptoasset transactions data are typically publicly available on a blockchain. Public availability of a complete history of different cryptoasset transaction data allows researchers to inspect various aspects of the cryptoasset market~\cite{nakamoto2008bitcoin}. Among the various cryptoassets, Bitcoin, Ethereum and XRP are the popular and large market cap cryptoassets. L.  Kristoufek studied the relationship of Bitcoin price with search queries on Google Trends and visit frequency on the Wikipedia page on Bitcoin~\cite{kristoufek2013bitcoin}. An early study~\cite{reid2013analysis} investigated the topological structure of the Bitcoin transaction network. D. Kondor {\em et. al.} demonstrated that the linear preferential attachment is the key mechanism for growth of the Bitcoin transaction network~\cite{kondor2014rich}. Ethereum transaction data has also been investigated and it is found that the transaction network exhibits power-law degree distribution, disassortativity, the absence of rich club phenomenon, and small world phenomenon ~\cite{guo2019graph, ferretti2020ethereum}. 
The cryptoasset transaction networks are growing and evolving with time. Application of principal component analysis on the structural change of Bitcoin transaction network shows a connection with Bitcoin price~\cite{kondor2014inferring}. It is also found that the out-degrees of the Bitcoin transaction network provides a connection to the price changes~\cite{vallarano2020bitcoin}. Recently, transaction data of crypto asset, XRP is also stuided. The structural properties, such as the heavy tail nature of degree distribution and triangular motifs are analyzed for XRP transaction netwrok~\cite{ikeda2022characterization}. Considering outgoing and incoming flows for both, XRP and 
Bitcoin transaction networks, the key nodes have been shown to be classified into three different groups~\cite{aoyama2022cryptoasset}. Remittance transactions recorded on the XRP ledger has also been studied recently~\cite{ikeda2022hodge}.

In this article, we focus on XRP, which was created by Ripple Labs in 2012. It is used for transferring value on the Ripple payment protocol and is also the native digital asset of the XRP ledger. Using the XRP transaction data, we have recently introduced a method of correlation tensor spectra of transaction network to detect price burst in XRP price~\cite{chakraborty2023projecting}.  The method of the correlation tensor is inspired by the cross correlation method in stock price time series~~\cite{laloux1999noise, plerou1999universal}. The theory of random matrices~\cite{sengupta1999distributions, mehta2004random, potters2020first} is crucial to understanding the structure of empirical correlation matrices. The method is very useful for separating noise from the signal by excluding the components that arises due to randomness. While this method is generally applied to price time series data, the correlation tensor method adapts this technique for transaction network snapshots.
\begin{figure}[h]
\includegraphics[width=0.78\textwidth]{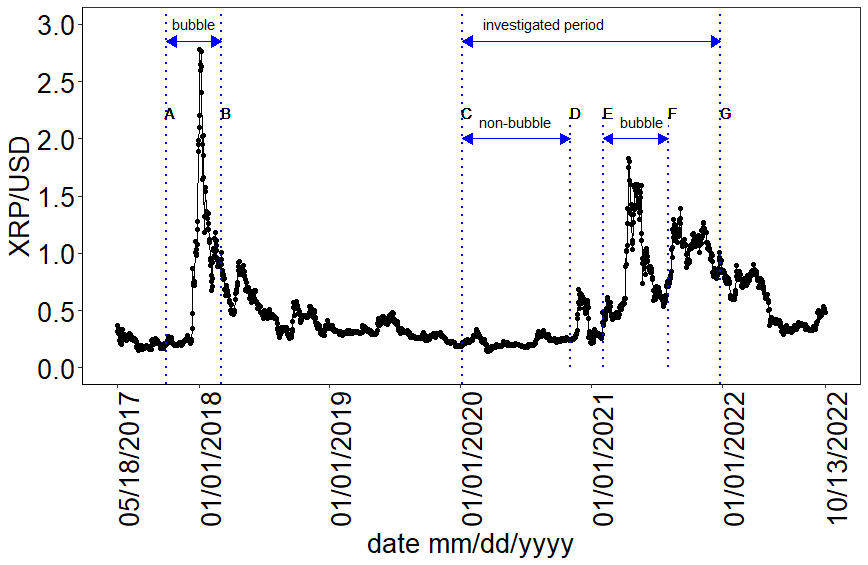}
\caption{{ The XRP/USD daily closing price is recorded from May 05, 2017 to October 13, 2022.}
The blue horizontal line segments between different pair of blue vertical lines represent different periods, which are explained in the main text.}
\label{fig1}
\end{figure}

The method of correlation tensor spectra was applied for the period of October 2017 to March 2018 which was the most significant bubble period of XRP price history~\cite{chakraborty2023projecting}. 
In this article, we study the robustness of correlation tensor spectra method over two years period between January 2020 to December 2021. We uncover the behaviour of correlation tensor spectra in both, bubble period and non bubble period.  The different periods for daily XRP price  are shown in figure~\ref{fig1}.  The periods $AB$, $CG$, $CD$ and $EF$ represent the period October 2, 2017 to March 4, 2018, January 6, 2020 to December 26, 2021, January 6, 2020 to November 1, 2020, and  February 1, 2021 to August 1, 2021 respectively.
The most significant bubble period for XRP price is observed in 2018, which is denoted by the period $AB$ within vertical lines A and B. In this article, we study the method correlation spectra on the period $CG$ within the vertical lines $C$ and $G$, which covers a period of $103$ weeks. We further delve into two sub periods: a non-bubble period $CD$ within vertical lines $C$ and $D$, and a bubble period $EF$ within the vertical lines $E$ and $F$. From the singular vectors of the empirical correlation tensor, we identify a set of small number of nodes that drives the XRP market.    

\begin{figure}[h]
\includegraphics[width=0.9\textwidth]{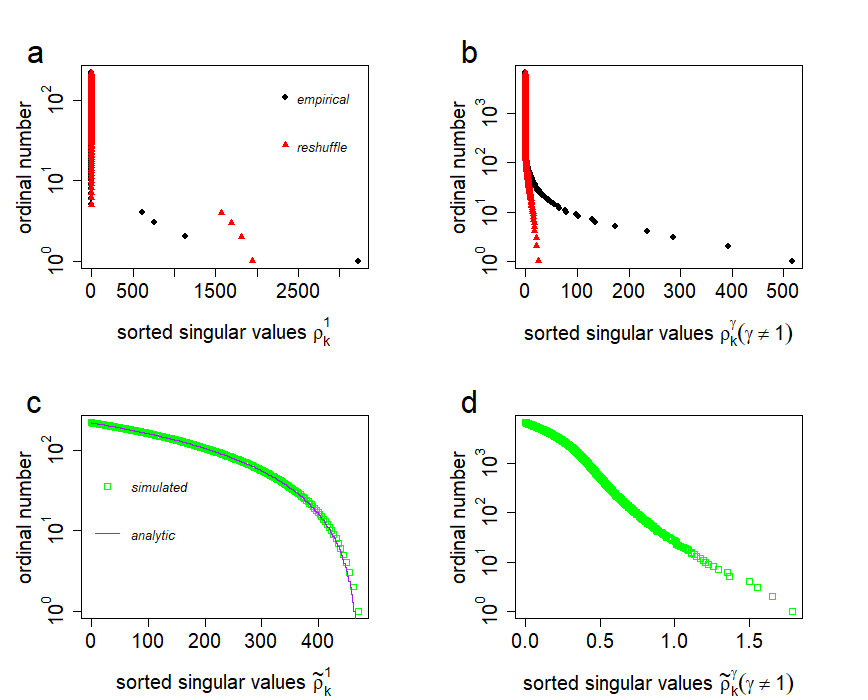}
\caption{{ Comparison of singular values for different correlation tensors for the the week January 06 -12, 2020.} 
(a) The plot shows the comparison of the singular values, $\rho_k^1$, for the empirical correlation tensor (black filled circle) and the reshuffled correlation tensor (red filled triangle) for all values of $k$. (b) The comparison of the singular values, $\rho_k^\gamma$, for the empirical correlation tensor and the reshuffled correlation tensor, considering all values of $k$ and $\gamma > 1$. (c) The simulated singular values, $\rho_k^1$, for the Gaussian random correlation tensor (green open square), along with the corresponding analytic curve following equation~\ref{eqn8} (solid purple line).  (d) The simulated singular values, $\rho_k^\gamma$, for the Gaussian random correlation tensor, considering all values of $k$ and $\gamma > 1$.}
\label{fig1b}
\end{figure}

\section{Materials and methods}

\subsection{Data}
We have collected the data from Ripple API. In this study, we mainly focus on the transaction data from January 06, 2020 to December 26, 2021, which is shown as the period $CG$ in figure~\ref{fig1}. The entire data is grouped into $103$ weeks. For each week, we construct a weekly weighed directed network of XRP transactions. The nodes of the networks are wallets. A link is formed from the source wallet to the destination wallet if there is at least one transaction between them. The link-weight represents the total transaction volume between the wallet for a week. Therefore, the link indicates the flow of XRP between wallets. The network statistics are shown in SI Text 1.  

\subsection{Network embedding}
We utilized the well-known node2vec~\cite{grover2016node2vec} algorithm to embed each of the weekly weighted directed networks into a $D$-dimensional space. In the node2vec algorithm, we have used the return parameter, $p = 1$ and in-out parameter, $q =1$, which represent the utilization of unbiased random walks. 
This results in a D-dimensional vector, denoted as $V_i^\alpha$, for every node in the networks. Here, we use $i$ and $j$ as node indices, and  $\alpha$ and $\beta$ as components of the vectors in the D-dimensional space. 

\subsection{Correlation tensor}
The method of the correlation tensor and its diagonalization using double Singular Value Decomposition (SVD) were introduced in~\cite{chakraborty2023projecting}. Here, we briefly discuss the method.  

In the weekly networks of XRP transactions, we identify $N$ nodes that carry out at least one transaction every week during the period under investigation. We refer to these nodes as regular nodes. Each regular node in the embedding space is represented by a time series of D-dimensional vectors, denoted $V_{i}^\alpha (t)$. Here, $i$ ranges from $1$ to $N$, $t$ ranges from $1$ to $T$, and $\alpha$ ranges from $1$ to $D$.

The correlation tensor between regular node components is given by:

\begin{equation}
M_{ij}^{\alpha\beta}(t) =\frac{1}{2\Delta T}\sum\limits_{t^\prime=t-\Delta T}^{t+\Delta T}\frac{[V_{i}^\alpha (t^\prime) - \overline{V_{i}^\alpha}][V_{j}^\beta (t^\prime) - \overline{ V_{j}^\beta}]}{\sigma_{V_i^\alpha} \sigma_{V_j^\beta}},
\label{eqn1}
\end{equation}
In this equation, we take the sum over five weekly networks at times $t^\prime = \{t-2, t-1, t, t+1, t+2\}$ with a time window of $(2 \Delta T + 1)$ with $\Delta T=2$ for our analysis. The values of $\overline{V_{i}^\alpha}$ and $\sigma_{V_i^\alpha}$ represent the mean and standard deviation of $V_{i}^\alpha$ over a time window of $(2 \Delta T + 1) = 5$ weekly networks at times $\{t-2, t-1, t, t+1, t+2\}$. It is important to note that a smaller value of $\Delta T$ results in more noise in the correlation tensor. However, we cannot choose a large value for $\Delta T$ as we are conducting a detailed temporal evolution of the networks. For this analysis we have chosen the dimension $D = 32$. The dependence of correlation tensor on window size $(2 \Delta T + 1)$ and dimension $D$ can be found in~\cite{chakraborty2023projecting}.

\subsection{Double singular value decomposition}
To determine the spectrum of the correlation tensor, we use a double SVD approach as follows:
First, we diagonalize $M_{ij}^{\alpha\beta}$ successively by a bi-unitary transformation, also known as SVD, in terms of the $(ij)$-index and then the $(\alpha\beta)$-index.
The first step involves expressing $M_{ij}^{\alpha\beta}$ as a sum of matrices, using the SVD method:
\begin{equation}
M_{ij}^{\alpha\beta} = \sum\limits_{k=1}^N L_{ik}\sigma_k^{\alpha\beta} R_{kj}.
\label{eqn2}
\end{equation}
The second step is to further decompose each singular value $\sigma_k^{\alpha\beta}$ as a sum of matrices, using SVD:
\begin{equation}
\sigma_k^{\alpha\beta} = \sum\limits_{\gamma=1}^D \mathcal{L}^{\alpha\gamma} \rho_k^\gamma \mathcal{R}^{\gamma\beta}.
\label{eqn3}
\end{equation}
Finally, we combine these steps to obtain the following expression for $M_{ij}^{\alpha\beta}$:
\begin{equation}
M_{ij}^{\alpha\beta} = \sum\limits_{k=1}^N \sum\limits_{\gamma=1}^D \rho_k^\gamma (L_{ik} R_{kj}) (\mathcal{L}^{\alpha\gamma} \mathcal{R}^{\gamma\beta}).
\label{eqn4}
\end{equation}
Here, $\rho_k^\gamma$ represents the $N \times D$ generalized singular values, which are real and positive due to the fact that $M$ is a real correlation tensor.

\section{Results}

\begin{figure}[h]
\includegraphics[width=0.98\textwidth]{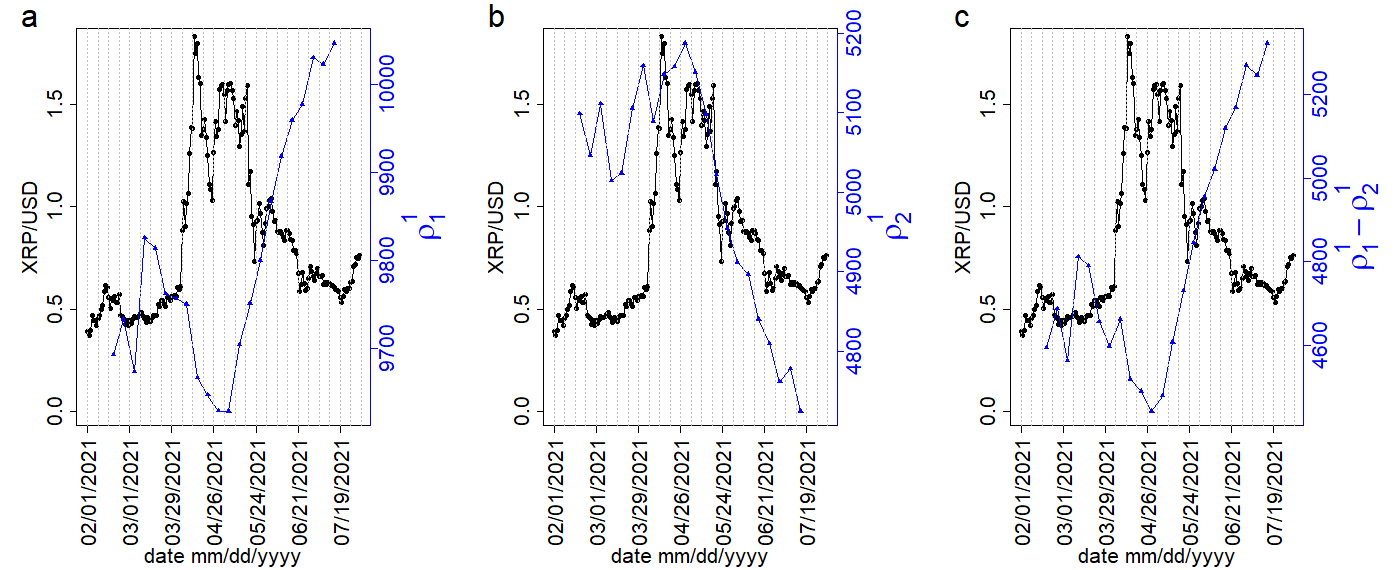}
\caption{{The comparison between the daily XRP/USD price with the singular values and spectral gap for the period, $EF$, February 01, 2021 to August 1, 2021.}
The black curves in the graph show the daily XRP/USD price, while the blue curves represent (a) the largest singular value $\rho_1^1$, (b) the second largest singular value $\rho_2^1$, and (c) the spectral gap  $(\rho_1^1-\rho_2^1)$ of correlation tensors for different weeks. The dotted grey vertical lines indicate the weekly windows.}
\label{fig3}
\end{figure}

We investigate the period,  $CG$, which spans $103$ weeks. It consists of $265$ regular nodes. Following equation~\ref{eqn1}, we calculate the correlation tensor between the components of regular nodes for different weeks. Note that with $103$ weekly networks, we get $99$ weekly correlation tensors following equation~\ref{eqn1}. A weekly correlation tensor has $N \times N \times D \times D$ elements. To get crucial information from the correlation tensor, we diagonalize it by double SVD as described in the method section. The double SVD is an extension of the SVD which is applied for a martix.  
Applying double SVD on the weekly correlation tensor $M_{ij}^{\alpha,\beta} (t)$, we get the singular values $\rho_k^\gamma (t)$.

\begin{figure}[h]
\includegraphics[width=0.98\textwidth]{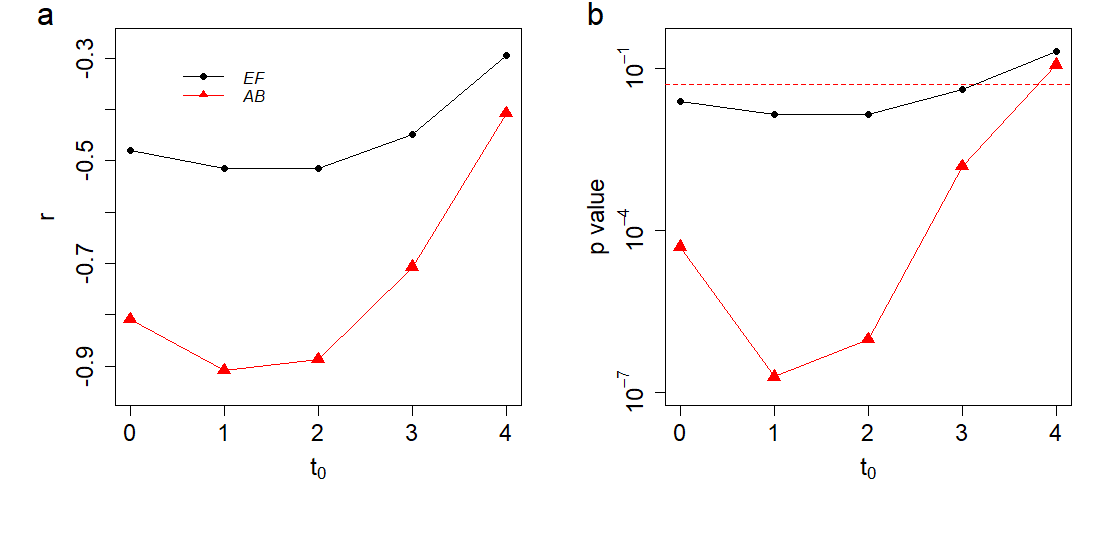}
\caption{{ The correlation between the weekly XRP/USD price $\overline{\rm XRP/USD}(t+t_0)$ and the largest singular value  $\rho_1^1(t)$ as a function of time lag $t_0$.}
(a) The Pearson correlation coefficient $r$ and (b) the associated p value are plotted with lag $t_0$.
The curves with black circles and red triangles represent the bubble periods $EF$ and $AB$ in figure~\ref{fig1} respectively.
The horizontal red dotted line in (b) indicates the significance label for p-values $< 0.05$.
}
\label{fig3b}
\end{figure}

 The significance of the empirical correlation tensor is measured by comparing it with the reshuffled correlation tensor.  To calculate the reshuffled correlation tensor, we reshuffle the components of the embedded regular node vector $v_i^\alpha$ with in the time window $(2\Delta T +1)$. Using the reshuffled embedded regular node vectors, we calculate the reshuffled correlation tensor following equation~\ref{eqn1}. We also calculate and simulate singular values of a Gaussian random correlation tensor using random matrix theory~\cite{sengupta1999distributions, edelman2005random, bouchaud2009financial, rudelson2010non, bryc2020singular} for comparison. The Gaussian correlation tensor elements $G_{ij}^{\alpha,\beta}$ are sampled from a Gaussian distribution with a mean of zero and a standard deviation of $\sigma_G=0.5$, where $(i,j = 1, \ldots, N)$ and $(\alpha, \beta = 1, \ldots, D)$.
We choose $\sigma_G=0.5$ to match the standard deviation of our empirical correlation tensor.
The probability distribution function form of the largest singular values of Gaussian random correlation tensor $(\tilde{\rho}_k^1)$ for all $k$ is given by 
\begin{equation}
P(\tilde{\rho}_k^1) = \frac{1}{\pi \sigma_G^2} \sqrt{(\tilde{\rho}_1^1)^2-(\tilde{\rho}_k^1)^2}, 
\label{eqn8}
\end{equation}
where $\tilde{\rho}_1^1=2\sigma_G D \sqrt{N}$ is the largest singular value for $k=1$. The derivation is shown in SI text 3.

We show the singular values $\rho_k^\gamma (t) $ of empirical, reshuffled and Gaussian random correlation tensor $M(t)$ for the week $t=$  January 06 -12, 2020 in figure~\ref{fig1b}.  Figure~\ref{fig1b}~(a) shows the singular values $\rho_k^\gamma $ for all $k \in [1,N]$ and $\gamma =1$ along with the singular values for the reshuffled correlation tensor. We observed that only the largest singular value $\rho_1^1$ for the empirical correlation tensor lies above the largest singular value for the  reshuffled correlation tensor. Similarly we show the comparison for other singular values $\rho_k^\gamma$ for all $k \in [1,2,3, \dots, N]$ and $\gamma \in [2,3,4, \dots, D]$. Here we observe that several singular values of the empirical correlation tensor lies above the largest of these singular values for the reshuffled correlation tensor. However, these singular values are much smaller than $\rho_1^1$. Therefore, these singular values have relatively less contribution to the correlation tensor.  We further compare the empirical singular values with the singular values of the Gaussian random correlation tensor, where the elements are taken from a normal distribution. We have calculated the singular values of the Gaussian random correlation tensor in equation~\ref{eqn8} using RMT.  
Figure~\ref{fig1b}~(c) shows the the simulated singular values, $\rho_k^\gamma$, of Gaussian correlation tensor fit nicely with the analytic curve given by equation~\ref{eqn8}. Figure~\ref{fig1b}~(d)  shows the spectrum, $\tilde{\rho}_k^\gamma$, of Gaussian random correlation tensor for all $k$ and $\gamma \in [2,3,4,\dots, D]$. It illustrates that these singular values are extremely small.
Here we observe that singular values for the Gaussian correlation tensors are much smaller than the singular values for the empirical correlation tensor and as well as for the reshuffled correlation tensor. It is observed that  the singular values of the reshuffled correlation tensor approaches to the singular values of the Gaussian correlation tensor when the time window $\Delta T$ becomes much larger than $N$ ~\cite{chakraborty2023projecting}. Although we have shown the results for the week,  January 06 -12, 2020 only, our results will remain qualitatively the same for any other week. 


\begin{figure}[h]
\includegraphics[width=0.98\textwidth]{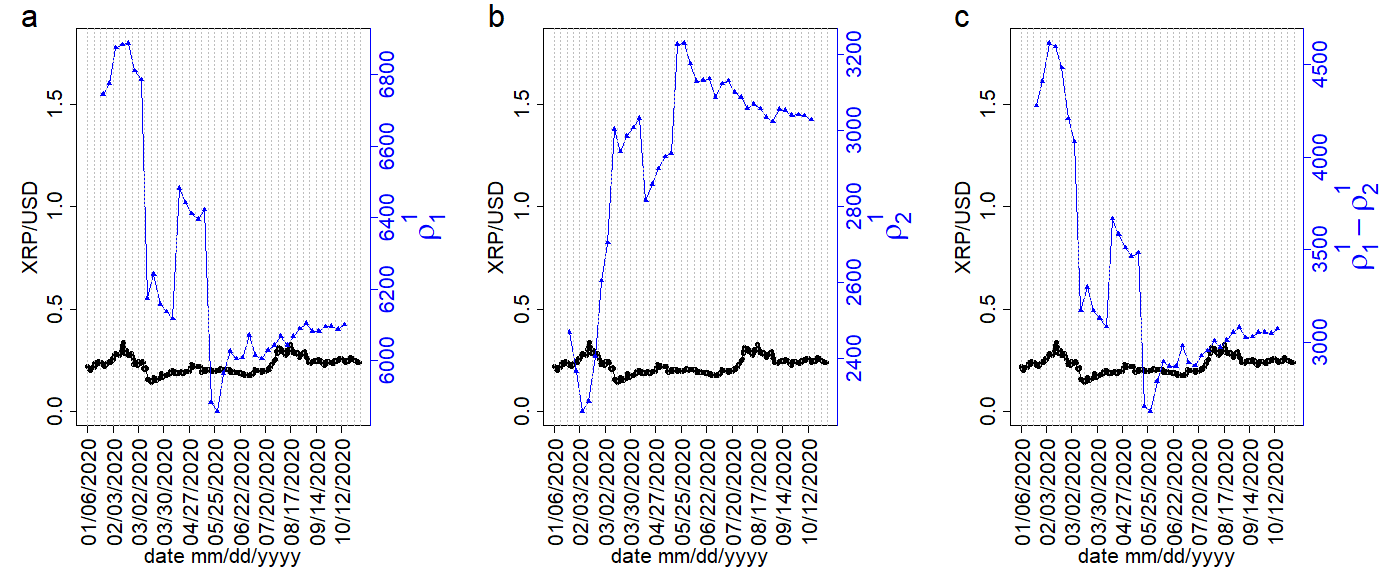}
\caption{{ The comparison between the daily XRP/USD price with the singular values and spectral gap for the period, $CD$ January 06, 2020 to November 01, 2020.}
The black curves in the graph show the daily XRP/USD price, while the blue curves represent (a) the largest singular value $\rho_1^1$, (b) the second largest singular value $\rho_2^1$, and (c) the spectral gap  $(\rho_1^1-\rho_2^1)$ of correlation tensors for different weeks. The dotted grey vertical lines indicate the weekly windows.}
\label{fig4}
\end{figure}

\begin{figure}[h!]
\includegraphics[width=0.98\textwidth]{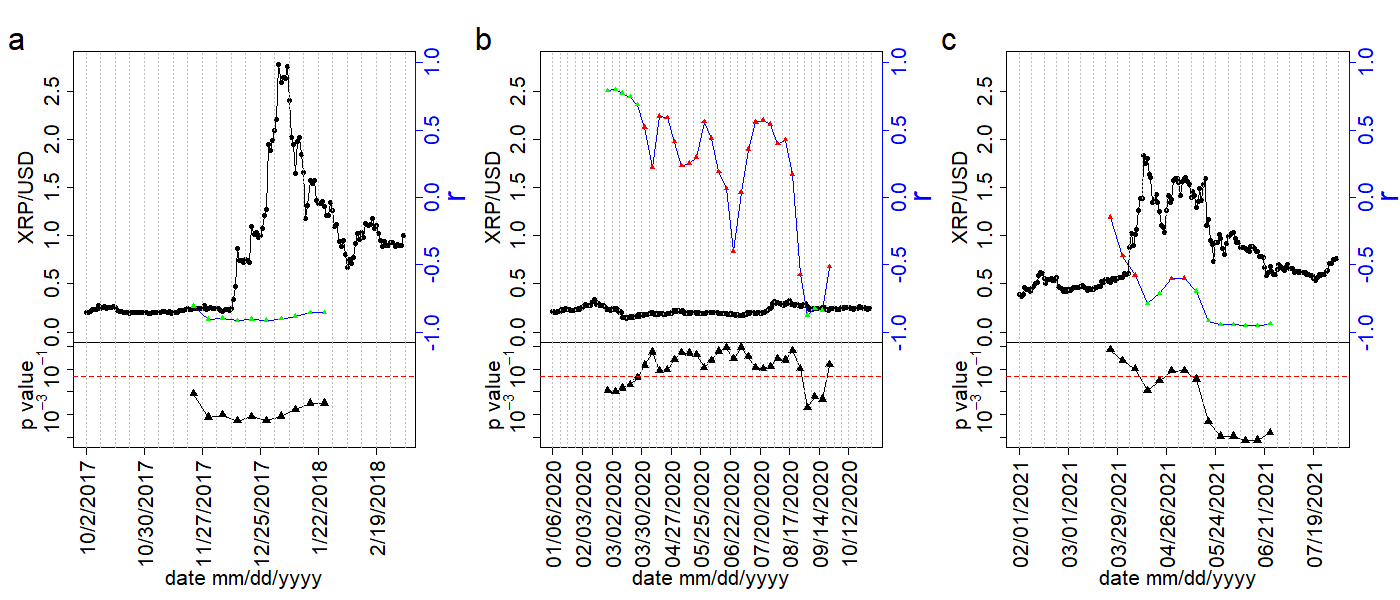}
\caption{ The comparison of daily XRP/USD price with the correlation $r(t)$ between the weekly XRP/USD price, $\overline{\rm XRP/USD}(t)$ and the largest singular value $\rho_1^1 (t-1)$ using a moving window of $9$ weeks for three different periods - (a) $AB$, October 2, 2017- March 4, 2018, (b) $CD$,  January 6, 2020 - November 1, 2020 and (c) $EF$, February 1, 2021- August 1, 2021.
The black curve represent daily XRP/USD closed price. The blue curve with green and red triangles represent correlation $r(t)$, where the green triangles indicate significant correlations (p-value $< 0.05$) and the red triangles indicate no significant correlations (p-value $> 0.05$)
The three lower panels show the p-values for the corresponding Pearson correlations. 
The dotted grey vertical lines represent the weekly windows. }
\label{fig6}
\end{figure}

The period $CG$ contains bubble and non bubble periods. To further explore how the relationship changes between the weekly XRP/USD price $\overline{\rm XRP/USD}(t)$  and the singular values $\rho_k^\gamma (t)$ during non-bubble and bubble periods, we separately study the following two sub-periods: January $06, 2020$ - November $01, 2020$ and February $01, 2021$ - August $01, 2021$, which is indicated as $CD$ and $EF$, respectively, in figure~\ref{fig1}. The weekly XRP/USD price, $\overline{\rm XRP/USD}(t)$ is calculated as the average daily XRP/USD price for the week. The analysis for the period $CG$ is given in SI Text 2.

The sub-period, $EF$, was a bubble period for XRP/USD price. In this period, we find that the weekly networks of XRP transactions has $753$ regular nodes. Taking into account these regular nodes, we calculate the weekly correlation tensors for this period. 
The two largest singular values and the spectral gap of the correlation tensors with the daily XRP prices are shown in figure~\ref{fig3}. We observe a strong anti-correlation for the weekly XRP/USD price $\overline{\rm XRP/USD}(t+1)$ with the largest singular value $\rho_k^\gamma(t)$ $ (r=-0.515$ and p-value$=0.014)$ and the spectral gap $(\rho_1^1(t)-\rho_2^1(t))$ $(r=-0.517$ and p-value $=0.014)$. A strong correlation is found between $\overline{\rm XRP/USD}(t+1)$ and the second largest singular value, $\rho_2^1(t)$ $(r=0.512$ and p-value $=0.015)$.

We further show the Pearson correlation between the weekly XRP price $\overline{\rm XRP/USD}(t+ t_0)$ and  the largest singular value $\rho_1^1(t)$ with different time lag in figure~\ref{fig3b} for the period $AB$ and $EF$ respectively. The anti correlation between weekly XRP price and the largest singular value is found to be maximum when the lag is either one week or two weeks. The correlation between $\overline{\rm XRP/USD}(t+3)$ and $\rho_1^1(t)$ is found  $r=-0.448$ and p-value $=0.041$ for the period $EF$. This is an indication that the largest singular value $\rho_1^1(t)$ can give an early warning for the XRP price burst, which is consistent with~\cite{chakraborty2023projecting}.

The sub-period, CD, indicates relatively steady values for XRP/USD price with a range of $0.30-0.15$. This period has 465 regular nodes for the weekly networks. Here also with these regular nodes, we calculate the weekly correlation tensors. The two largest singular values and spectral gap for the weekly correlation tensors are shown together with daily XRP/USD price in figure~\ref{fig4}. We observe that the weekly XRP/USD price $\overline{\rm XRP/USD}(t+1)$ has no significant correlation with the largest singular value $(r=0.193$ and p-value $=0.238)$ and the spectral gap $(r=0.259$ and p-value $=0.111)$. It only shows a weak anti-correlation between $\overline{\rm XRP/USD}(t+1)$ and the second largest singular value $(r=-0.333$ and p-value $=0.039)$. Therefore, we observe a distinct behaviour of the weekly XRP price and the largest singular value for the period $CD$ in stark contrast to its strong anti correlation behaviour during bubble periods $AB$ and $EF$.


\begin{figure}[h]
\includegraphics[width=0.98\textwidth]{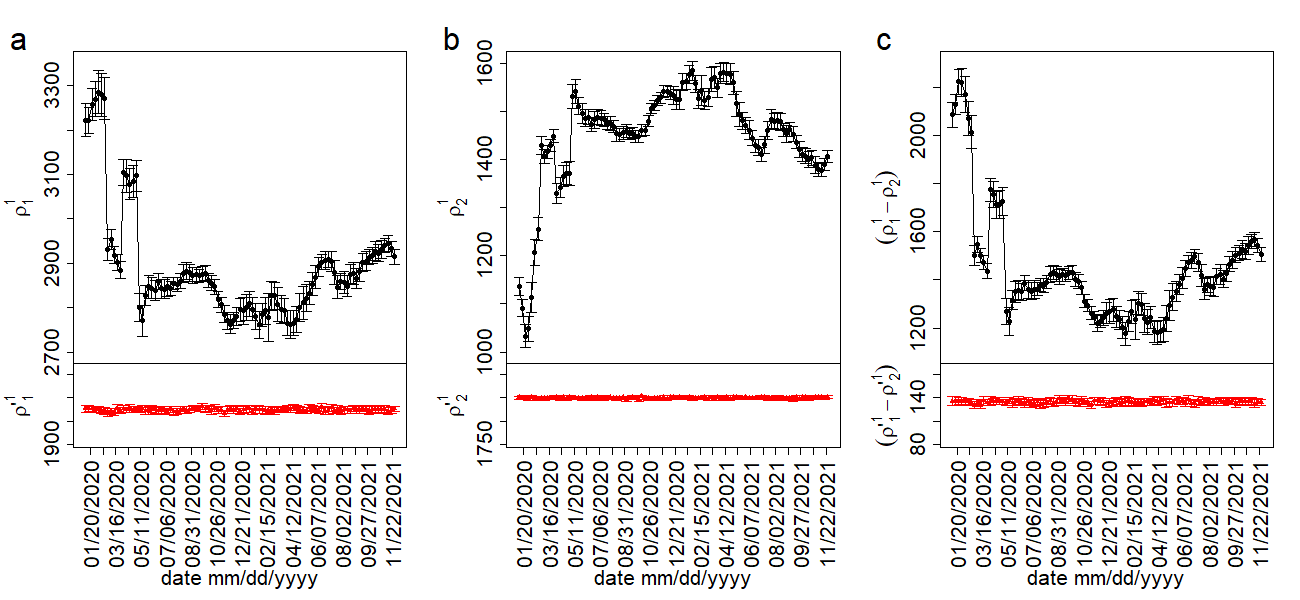}
\caption{{ The variation of singular values and spectral gaps of empirical correlation tensor and reshuffled correlation tensor for different weeks for the period $CG$. }
  (a) Variation of  the largest singular value for empirical $\rho_1^1$ and reshuffled $\rho_1^{\prime1}$ correlation tensors.   (b) Variation of the second largest singular value for empirical $\rho_2^1$, reshuffled $\rho_2^{\prime1}$ correlation tensors. (c) Variation of spectral gap for empirical  $(\rho_1^1 - \rho_2^1)$ and reshuffled  $(\rho_1^{\prime1} - \rho_2^{\prime1})$ correlation tensors. The error bars on the graphs represent the standard deviation. The data is an average of $40$ different network embeddings. 
  }
\label{fig7}
\end{figure}

To show the temporal relationship between the XRP price and tensor correlation spectra,
we calculate the Pearson correlation between weekly XRP/USD price $\overline{\rm XRP/USD}(t)$ and the largest singular value $\rho_1^1 (t-1)$ using a moving time window of length $9$ weeks. 
The Pearson's correlation coefficient $r(t)$ is given by  
\begin{equation}
r(t) =\frac{1}{2\Delta \tau}\sum\limits_{t^\prime=t-\Delta \tau}^{t+\Delta \tau}\frac{[\overline{\rm XRP/USD}(t^\prime) - \langle \overline{\rm XRP/USD}\rangle][\rho_1^1(t^\prime -1) - \langle \rho_1^1 \rangle]}{\sigma_{\overline{\rm XRP/USD}} \sigma_{\rho_1^1}},
\label{eqn9}
\end{equation}
where we have taken $\Delta \tau = 4$. The $\sigma$ and $\langle \cdot \rangle$ represent standard deviation and mean of the quantities with in the time window $(2 \Delta \tau + 1)$ respectively. 
We  investigate the temporal correlation separately for three different periods, AB, CD and EF.
The number of regular nodes for the three periods are $71$, $465$ and $753$ respectively. The temporal variation of the correlation $r(t)$ between the weekly XRP prices $\overline{\rm XRP/USD}(t)$ and the largest singular values $\rho_1^1 (t-1)$ of the weekly correlation tensors is shown along with the  daily XRP/USD price in figure~\ref{fig6}. It is observed that the anti correlation is the strongest and significant during the period $AB$ and is stronger and significant during the period $EF$. The anti correlation mostly non-significant during the period $CD$. It reflects the fact that the formation of a large bubble in the XRP price is indicated by a strong anti correlation with the largest singular value.

The significance of the evolution of the singular values of the empirical correlation tensor is measured by comparing it with the singular values of the reshuffled correlation tensor. 
The two largest singular values and the spectral gap of the empirical correlation tensor for the period $CG$ are compared to those of the reshuffled correlation tensor in figure~\ref{fig7}.
It shows that only the largest singular value of the empirical correlation tensor is larger than its reshuffled counterpart. Moreover, we observe that while the singular values for the reshuffled correlation tensor remain approximately constant, those for empirical correlation tensor exhibit non-trivial variation over the investigated time period. 

\begin{figure}[h]
\centering
\includegraphics[width=0.75\textwidth]{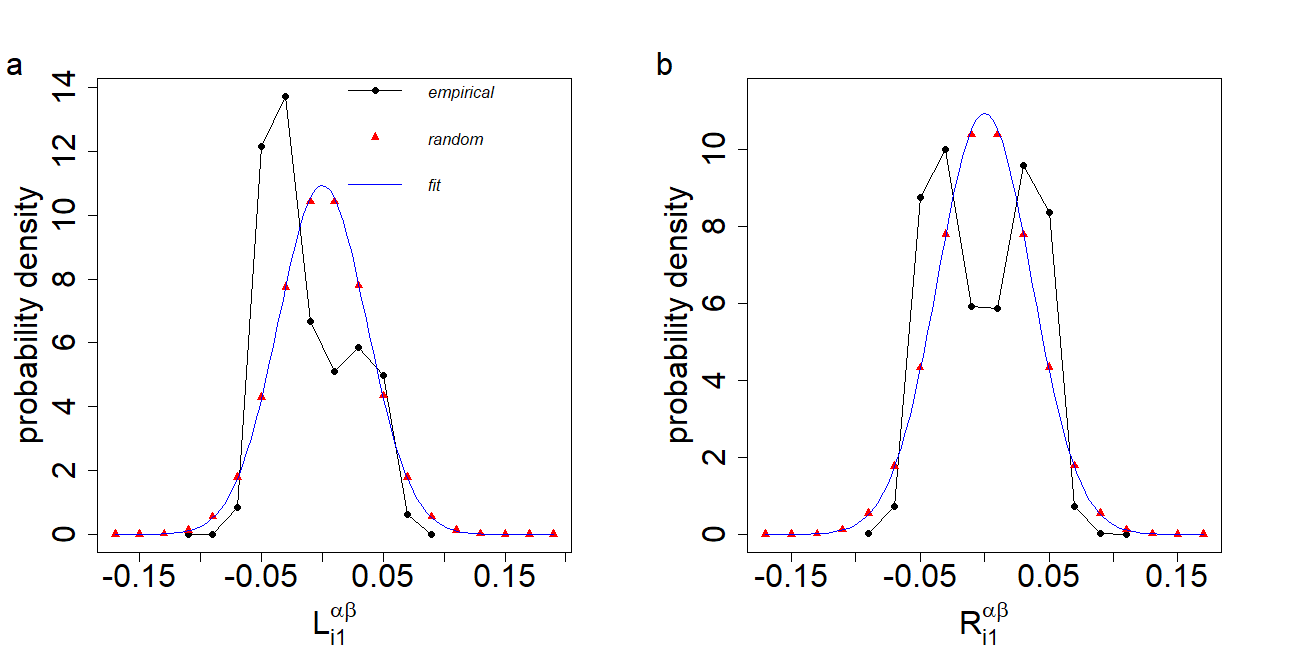}
\caption{The comparison of the distribution for the components of (a) left singular vectors  $L_{i1}^{\alpha,\beta}$ and (b) right singular vectors  $R_{i1}^{\alpha,\beta}$ between Gaussian random correlation tensor and empirical correlation tensor. Here $i \in [1,2,3, \dots, N]$, and $\alpha, \beta \in [1,2,3, \dots, D]$. The distributions for the components of  $L_{i1}^{\alpha,\beta}$ and $R_{i1}^{\alpha,\beta}$ fits nicely with  normal distribution for the Gaussian random correlation tensor. The mean and standard deviation for the fitted normal distributions are taken $6.0 \times 10^{-5}$, $0.036$ for  $L_{i1}^{\alpha,\beta}$ and $5.6 \times 10^{-5}$, $0.036$ for $R_{i1}^{\alpha,\beta}$ respectively. 
The empirical period is taken during the week April $5-11$, $2021$, in the bubble period. 
}
\label{f2}
\end{figure}

Up to this point, our focus has primarily centered around delving into the characteristics of singular values. Moving forward, our attention turns to a comprehensive exploration of the singular vectors.  
We provide a comparison of the distribution of the components of singular vectors for the empirical correlation tensor with the Gaussian random correlation tensor. We mainly focus on the largest left singular vectors $L_{i1}^{\alpha,\beta}$ and  the largest right singular vectors $R_{i1}^{\alpha,\beta}$ to identify nodes, $i$, that play key role in the transaction networks. 
The comparison of the distributions for the singular vectors' components between the Gaussian random correlation tensor and the empirical correlation tensor is shown in figure~\ref{f2}.
We can observe differences in the nature of the peaks in the distributions. The distribution for both the left and right largest singular vector components for the random Gaussian correlation tensor follows the Gaussian distribution. For the empirical correlation tensor it is bimodal and thus far from Gaussian in nature. 

\begin{figure}[h]
\centering
\includegraphics[width=0.98\textwidth]{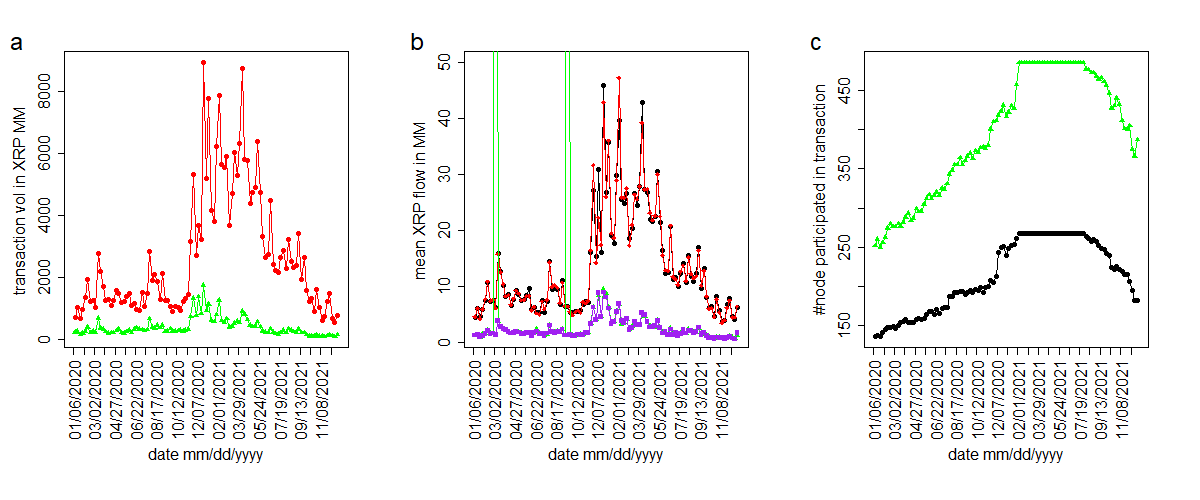}
\caption{ Comparative analysis of XRP Transaction volumes, flows, and numbers between the driver set of nodes and the remaining regular set of nodes for the largest left singular vectors, $L_{i1}^{\alpha,\beta}$ ($t=$ April $5-11, 2021$) over the period $CG$.
(a) Comparison of the weekly transaction XRP volumes (in millions) between the driver set of nodes (red) and the remaining regular set of nodes (green). (b) Comparison of mean inflow and outflow of XRP (in millions) between the driver set of nodes (black and red) and the remaining regular set of nodes (green and purple).
(c) Number of nodes present in the weekly networks for the driver set of nodes (black circle) and the remaining regular set of nodes (green triangle). 
}
\label{f4}
\end{figure}

Nodes with larger values of $L_{i1}^{\alpha,\beta}$ hold more significance in the correlation tensor. We have observed that the distribution of $L_{i1}^{\alpha,\beta}$ falls within the range of $(-0.1, 0.1)$.
To determine which node indices are overrepresented among the large values of $L_{i1}^{\alpha,\beta}$, we have set a threshold of 0.05. Any value of $|L_{i1}^{\alpha,\beta}|$ exceeding this threshold is considered significant. We then calculate the total count, denoted by $N_c$, of $L_{i1}^{\alpha,\beta}$ values that surpass 0.05.
The expected frequency of occurrence for a particular node index $j$ within $N_c$ is evaluated by dividing $N_c$ by the total number of nodes, denoted by $N$. We consider a node index to be overrepresented in $N_c$ if its frequency surpasses $(N_c/N + 10)$. This threshold is deliberately set slightly higher than the expected frequency by random chance $N_c/N$.
Similarly, we identify the node indices that are overrepresented in the range $L_{i1}^{\alpha,\beta} < -0.05$.
For the period (bubble period), April 5-11, 2021, our analysis reveals that there are 112 node indices that are overrepresented for $L_{i1}^{\alpha,\beta} > 0.05$, while 189 node indices are overrepresented for $L_{i1}^{\alpha,\beta} < -0.05$. Combining these two sets, we find a total of 268 unique node indices that are considered important based on our criteria.
To gain further insights, we compare the total transaction volume, mean inflow and outflow of XRP for these 268 important nodes (referred to as the ``driver set") with the remaining regular nodes, which amount to 485 in total (753 nodes minus the 268 driver set nodes). This comparison is illustrated in figure~\ref{f4}. In figure~\ref{f4}~(a), we observe that the total transaction volume between the driver set of nodes increases during the bubble period, however such change in transaction volume is not detected for the remaining set of nodes. Notably, the mean inflow and outflow exhibit similar variations for both sets of nodes. However, what distinguishes the driver set is a noticeable jump in the mean inflow and outflow during the bubble period, suggesting a distinct behavior among these nodes [figure~\ref{f4}~(b)]. We also observe two sudden peaks in the mean inflow of remaining set of nodes during the week, March 16 - 22, 2020 and September 28 - October 04, 2020 respectively due to a large inflow $(~ 10^{11}$ XRP$)$ to a distinct wallet.  
Figure~\ref{f4}~(c) shows the temporal variation for total number of these node sets in the weekly transaction networks. 

Similarly, we conducted a similar analysis for the largest right singular vector, denoted as $R_{i1}^{\alpha,\beta}$, which yielded comparable results. For the bubble period of April 5-11, 2021, we identified 159 node indices that were overrepresented for $R_{i1}^{\alpha,\beta} > 0.05$, while 167 node indices were overrepresented for $R_{i1}^{\alpha,\beta} < 0.05$. Combining these two sets, we found a total of 247 unique node indices that met our criteria.

\begin{figure}[h]
\centering
\includegraphics[width=0.98\textwidth]{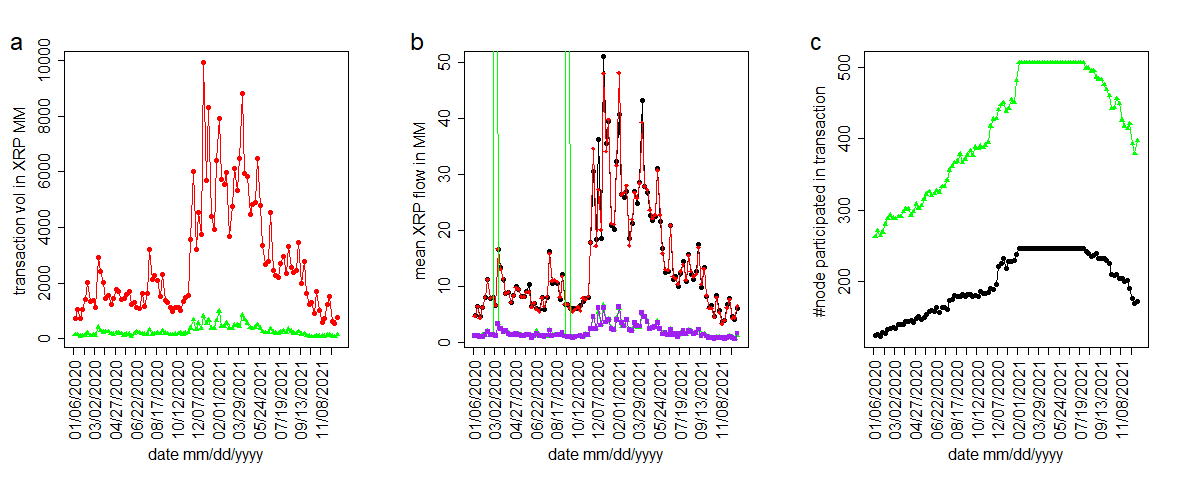}
\caption{Comparative analysis of XRP Transaction volumes, flows, and numbers between the driver set of nodes and the remaining regular set of nodes for the largest right singular vectors, $R_{i1}^{\alpha,\beta}$ ($t=$ April $5-11, 2021$) over the period $CG$.
(a) Comparison of the weekly transaction XRP volumes (in millions) between the driver set of nodes (red) and the remaining regular set of nodes (green). (b) Comparison of mean inflow and outflow of XRP (in millions) between the driver set of nodes (black and red) and the remaining regular set of nodes (green and purple).
(c) Number of nodes present in the weekly networks for the driver set of nodes (black circle) and the remaining regular set of nodes (green triangle).
}
\label{f5}
\end{figure}

To gain further insights, we compare the total transaction volume, the mean inflow and outflow of XRP for these 247 driver set nodes with the remaining regular nodes, which amounted to 506 in total (753 nodes minus the 247 driver set nodes). This comparison is depicted in figure~\ref{f5}.

Taking into account both the left and right largest singular vectors, we identified a total of 313 unique node indices in the driver set. These findings suggest that these particular nodes exhibit noteworthy behavior and warrant closer examination in the context of our analysis.

\section{Conclusion}
The analysis of time series data by cross correlation matrix~\cite{laloux1999noise,plerou1999universal} provides useful information. In the cross correlation method, the correlation is measured for different time dependent variables. such as stock price data, foreign exchange rate data or even medical data. However, in this study, we have used snapshots of weekly weighted directed XRP transaction networks. To measure the correlations between the regular nodes we have considered their embedded vectors. Correlations between the components of the regular nodes are represented as the correlation tensor. A double SVD has been used to get the singular values of the correlation tensor.   

We have studied the temporal variation of correlation between the XRP price and correlation tensor spectra of transaction networks. It is observed that the dependence between the XRP price and the largest singular value of the correlation tensor changes depending on the market situation. For the non-bubble period, we found that there is no significant correlation between the XRP price and the largest singular value. In stark contrast, for the bubble period, we observe a strong anti-correlation between the XRP price and the largest singular value. Moreover, the significance of the empirical singular values is shown by
a comparison with the singular values of the reshuffled correlation tensor. We have also provided an theoretical expression for the singular values of the Gaussian random correlation tensor using random matrix theory.  We also showed how the distributions for the components of singular vector deviate from the normal distribution which is followed in case of Gaussian random correlation tensor. From the singular vectors, we identified a small subset of nodes that drive the XRP market during the bubble period.



\section*{Acknowledgements}
We thank the members of the Kyoto University - RIKEN Blockchain Study Group for insightful discussions. YI acknowledges the financial support received from the Ripple Impact Fund (Grant Number: 2022-247584).

\newpage
	\section*{Supplementary Information}

	\renewcommand{\figurename}{SI Fig.}
	 \setcounter{table}{0}
        \renewcommand{\thetable}{S\arabic{table}}%
        \setcounter{figure}{0}
        \renewcommand{\thefigure}{S\arabic{figure}}%
        \setcounter{equation}{0}
        \renewcommand{\theequation}{S\arabic{equation}}%
\subsection*{SI Text 1: The weekly XRP transaction statistics}
\begin{figure}[h]
\centering
\includegraphics[width=0.98\textwidth]{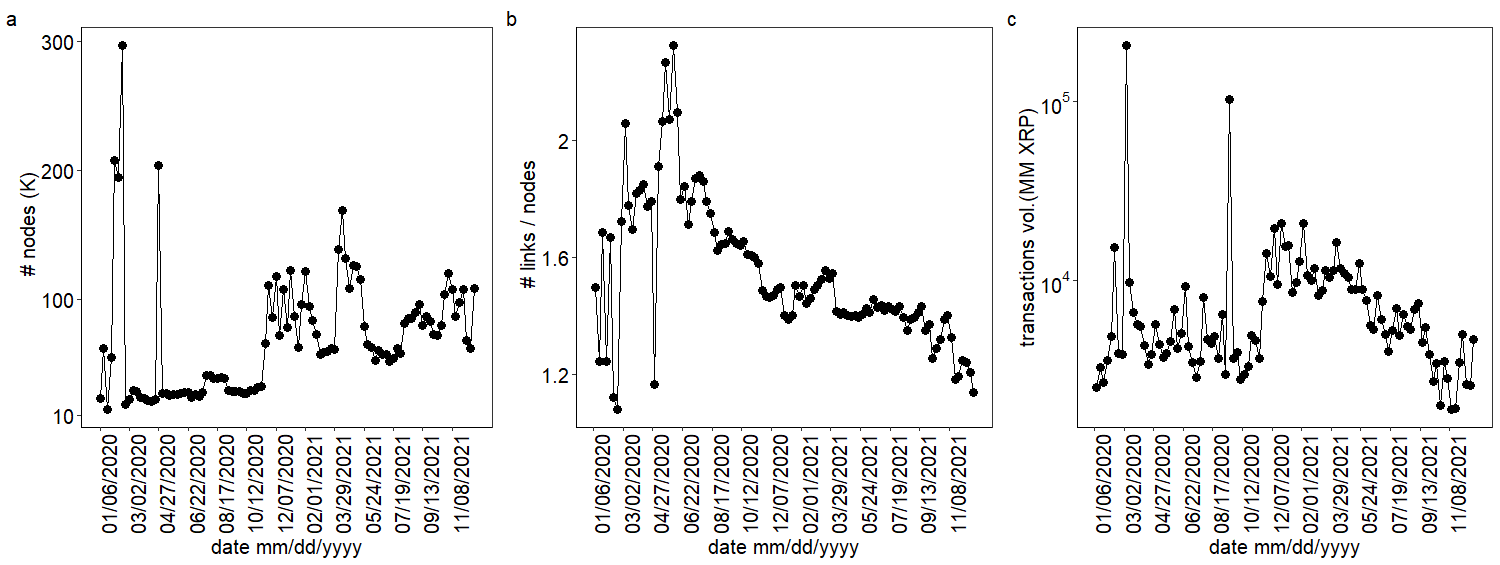}
\caption{ For each weekly network, the following variables exhibit variation: (a) the total number of nodes in thousands, (b) the number of links per node, and (c) the total transaction volume in millions of XRP.}
\label{Figs1}       
\end{figure}

The period, January 6, 2020 to December 26, 2021 covers 103 weeks of XRP transactions data. Here, we calculate the network statistics of each weekly weighted directed networks. We show the variation of number of nodes, links per nodes and total XRP transaction volume for each weekly XRP transaction network in figure.~\ref{Figs1}. 

\subsection*{SI Text 2: The temporal variation of the largest two singular and spectral gap with XRP/USD price for the period CG, January 06, 2020 to December 26, 2021}

\begin{figure}[h]
\includegraphics[width=0.98\textwidth]{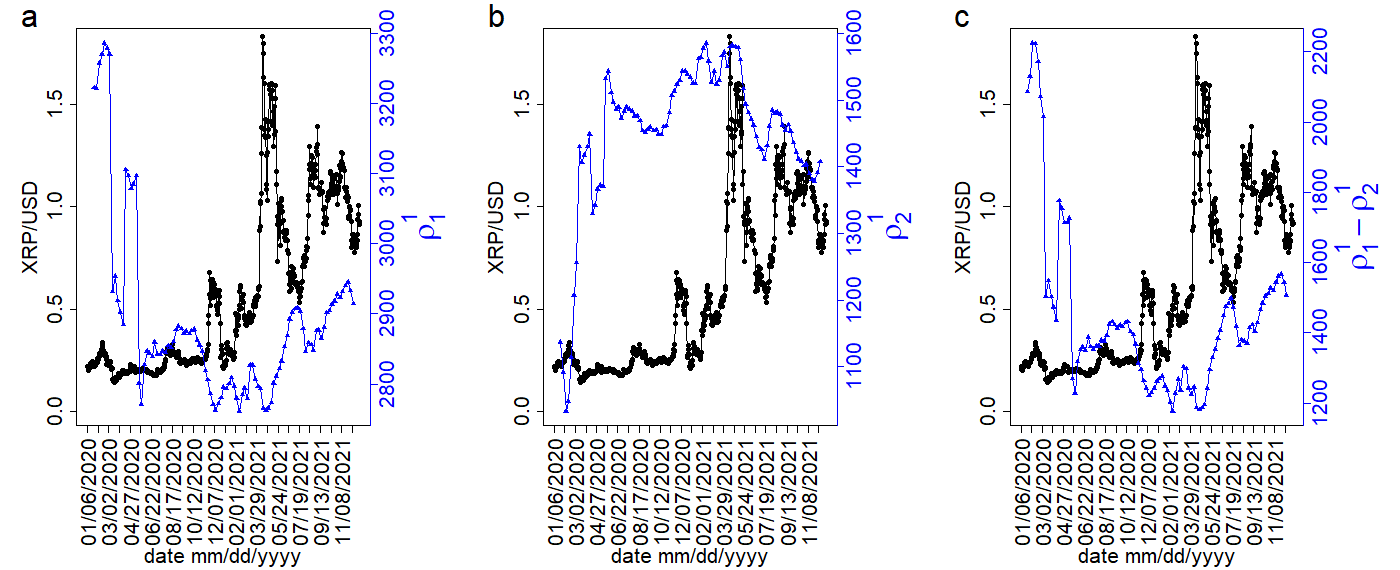}
\caption{{ The comparison between the daily XRP/USD price with the singular values and spectral gap for the period, $CG$, January 06, 2020 to December 26, 2021.}
The black curves in the graph show the daily XRP/USD price, while the blue curves represent (a) the largest singular value $\rho_1^1$, (b) the second largest singular value $\rho_2^1$, and (c) the spectral gap  $(\rho_1^1-\rho_2^1)$ of correlation tensors for different weeks.}
\label{FigS2}
\end{figure}

In figure~\ref{FigS2}, we compare the daily XRP/USD price with the largest two singular values and the spectral gap of the correlation tensor.
We find a significant but weak anti-correlation between the weekly XRP/USD price $\overline{\rm XRP/USD}(t+1)$ and the largest singular value $\rho_1^1(t)$. The Pearson correlation coefficient is found to be $r=-0.273$ with a p-value of $0.006$ between these two quantities. We also observe that the weekly XRP/USD price $\overline{\rm XRP/USD}(t+1)$ correlates weakly with the second largest singular value $\rho_2^1(t)$ $(r=0.243,$ and p-value $=0.015)$ and a weak anti-correlation with the spectral gap $(\rho_1^1(t)-\rho_2^1(t))$ $(r=-0.262,$ and p-value$=0.009)$.

\begin{figure}[t]
\includegraphics[width=0.78\textwidth]{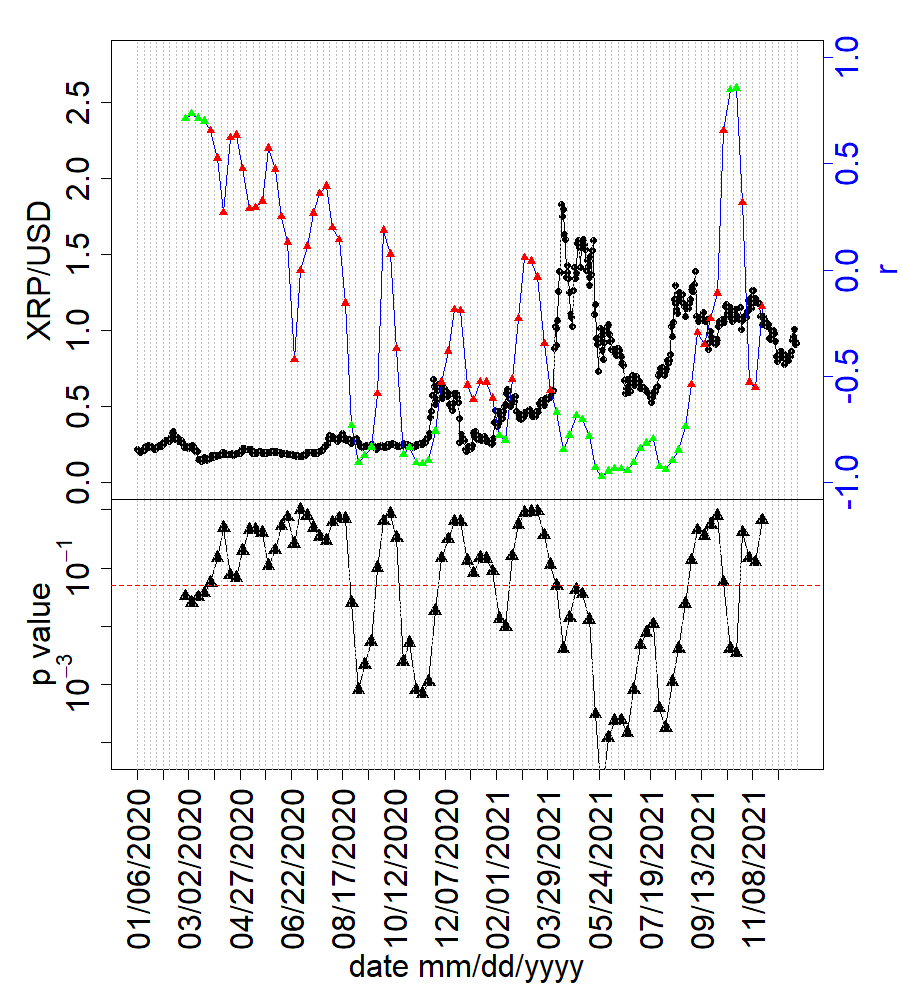}
\caption{ The comparison of daily XRP/USD price with the correlation $r(t)$ between weekyly XRP/USD price  and the largest singular value $\rho_1^1$ using a moving window of $9$ weeks. The black curve represent daily XRP/USD closed price. The blue curve with green and red triangles represent correlation $r(t)$, where the green triangles indicate significant correlations (p-value $< 0.05$) and the red triangles indicate not significant correlations (p-value $> 0.05$)
The lower panel shows the p-values for the corresponding Pearson correlations. 
The dotted grey vertical lines represent the weekly windows. }
\label{FigS3}
\end{figure}

To show the temporal relationship between the XRP price and correlation spectra,
we calculate the Pearson correlation between weekly XRP/USD price $\overline{\rm XRP/USD}(t)$ and the largest singular value $\rho_1^1 (t-1)$ using a moving time window of length $9$ weeks. Here, the Pearson's correlation coefficient $r(t)$ represents the middle point of the moving time window. 
A comparison between the Pearson correlation coefficients with daily XRP/USD price is shown in figure~\ref{FigS3}. It shows that during the bubble period, the largest singular values $\rho_1^1 (t-1)$ exhibit a strong anti correlation with  $\overline{\rm XRP/USD}(t)$. Furthermore, it shows that the largest singular values $\rho_1^1 (t-1)$ do not have any significant correlation with $\overline{\rm XRP/USD}(t)$ during the non-bubble period.

\subsection*{SI Text 3: Double singular value decomposition of Gaussian random correlation tensor}
 We also calculate and simulate the singular values using random matrix theory~\cite{sengupta1999distributions, edelman2005random, bouchaud2009financial, rudelson2010non, bryc2020singular} for a comparison. We consider a random Gaussian correlation tensor $G$. 
The tensor elements $G_{ij}^{\alpha,\beta}$ are sampled from a Gaussian distribution with a mean of zero and a standard deviation of $\sigma_G,$ where $(i,j = 1, \ldots, N)$ and $(\alpha, \beta = 1, \ldots, D)$.

A singular value decomposition of $G_{ij}^{\alpha,\beta}$ for fixed $\alpha,\beta $ gives
\begin{equation}
   G_{ij}^{\alpha,\beta} = \sum\limits_{k=1}^N \tilde{L}_{ik} \tilde{\sigma}_k^{\alpha\beta} \tilde{R}_{kj},
   \label{eqnS1}
\end{equation}
The distribution for $\tilde{\sigma}_k^{\alpha\beta}$ can be analytically calculated using the method described in~\cite{sengupta1999distributions}.
The probability distribution $P(\tilde{\sigma}_k^{\alpha\beta})$ for the singular values $\tilde{\sigma}_k^{\alpha\beta}$ for all $k$ with fixed $\alpha,\beta $ is given by
\begin{equation}
    P(\tilde{\sigma}_k^{\alpha\beta}) = \frac{1}{\pi \tilde{\sigma}_k^{\alpha\beta} \sigma_G^2}\sqrt{\{(\tilde{\sigma}_1^{\alpha\beta})^2 -(\tilde{\sigma}_k^{\alpha\beta})^2\}\{(\tilde{\sigma}_k^{\alpha\beta})^2 -(\tilde{\sigma}_N^{\alpha\beta})^2\}},
    \label{eqnS2}
\end{equation}
where $\tilde{\sigma}_1^{\alpha\beta} = 2 \sigma_G \sqrt{N} $ is the largest singular value and $\tilde{\sigma}_N^{\alpha\beta} \sim 0$ is the smallest singular value. Substituting $\tilde{\sigma}_N^{\alpha\beta} = 0 $ in equation~\ref{eqnS2}, it becomes  

\begin{equation}
    P(\tilde{\sigma}_k^{\alpha\beta}) = \frac{1}{\pi \sigma_G^2}\sqrt{(\tilde{\sigma}_1^{\alpha\beta})^2 -(\tilde{\sigma}_k^{\alpha\beta})^2}.
    \label{eqnS3}
\end{equation}

The distribution given in equation~\ref{eqnS3} remains unchanged when considering all $\alpha$ and $\beta$. This is due to the fact that for every set of $\alpha,\beta $ the elements of $G_{ij}^{\alpha,\beta}$ are sampled from the same Gaussian distribution.  

Now we consider a further singular value decomposition of $\tilde{\sigma}_k^{\alpha\beta}$ for a fixed k and all $\alpha,\beta $ as given by the following equation: 

\begin{equation}
\tilde{\sigma}_k^{\alpha\beta} = \sum\limits_{\gamma=1}^D \tilde{\mathcal{L}}^{\alpha\gamma} \tilde{\rho}_k^\gamma \tilde{\mathcal{R}}^{\gamma\beta}.
\label{eqnS4}
\end{equation}
The elements of $\tilde{\sigma}_k^{\alpha\beta}$ for a fixed k will have a peaked distribution around a fixed value. For example, $\tilde{\sigma}_1^{\alpha\beta}$ will have a peaked distribution around $2 \sigma_G \sqrt{N}$. 
The value of $\tilde{\rho}_k^1$ will be $\tilde{\sigma}_k^{\alpha\beta} D$ ~\cite{rudelson2010non, bryc2020singular}, while the other $\tilde{\rho}_k^\gamma$ values for $\gamma \neq 1$ will be very close to zero. Therefore, the shape of the distribution of the $\tilde{\rho}_k^1$  will remain same as given by equation~\ref{eqnS3}.
Therefore, the probability distribution function form of the largest singular values $(\tilde{\rho}_k^1)$ for all $k$ is given by 
\begin{equation}
P(\tilde{\rho}_k^1) = \frac{1}{\pi \sigma_G^2} \sqrt{(\tilde{\rho}_1^1)^2-(\tilde{\rho}_k^1)^2}, 
\label{eqnS5}
\end{equation}
where $\tilde{\rho}_1^1=2\sigma_G D \sqrt{N}$ is the largest singular value for $k=1$.



\end{document}